# Comprehensive transient-state study for CARMENES NIR high-thermal stability

Santiago Becerril [a*], Miguel A. Sánchez [a], M. C. Cárdenas [a], Ovidio Rabaza [a], Alejandro Ramón [a], Miguel Abril [a], Luis P. Costillo [a], Rafael Morales [a], Alicia Rodríguez [a], Pedro J. Amado [a] and the international CARMENES team

[a] Instituto de Astrofísica de Andalucía (CSIC), C/Glorieta de la Astronomía s/n, Granada, Spain, +0034 958 121 311

**ABSTRACT**

CARMENES has been proposed as a next-generation instrument for the 3.5m Calar Alto Telescope. Its objective is finding habitable exoplanets around M dwarfs through radial velocity measurements (m/s level) in the near-infrared. Consequently, the NIR spectrograph is highly constraint regarding thermal/mechanical requirements. Indeed, the requirements used for the present study limit the thermal stability to ±0.01K (within year period) over a working temperature of 243K in order to minimise radial velocity drifts. This can be achieved by implementing a solution based on several temperature-controlled rooms (TCR), whose smallest room encloses the vacuum vessel which houses the spectrograph's optomechanics.

Nevertheless, several options have been taken into account to minimise the complexity of the thermal design: 1) Large thermal inertia of the system, where, given a thermal instability of the environment (typically, ±0.1K), the opto-mechanical system remains stable within ±0.01K in the long run; 2) Environment thermal control, where thermal stability is ensured by controlling the temperature of the environment surrounding the vacuum vessel.

The present article also includes the comprehensive transient-state thermal analyses which have been implemented in order to make the best choice, as well as to give important inputs for the thermal layout of the instrument.

## 1. INTRODUCTION

CARMENES has been proposed as a next-generation instrument for the 3.5m Calar Alto Telescope. Its objective is finding habitable exoplanets around M dwarfs through radial velocity measurements in the near-infrared (within 1m/s accuracy). The instrument is to be mounted to either the Prime Focus or the Cassegrain Focus.

The scope of this paper is mainly constraint to the NIR spectrograph, the mechanical hardware attached to the Focal Plane not being object of the study. In order to obtain 1m/s accuracy on the radial velocity measurements, high degree of thermal/mechanical stability in the spectrograph is required. That is why the present concept of the instrument is based on two optical fibers which carry the light of both the object and the calibration unit to a spectrograph which is located in a stable, controlled environment.

According the tight thermal/mechanical requirements, this paper describes the thermal design concept which has been presented to the Conceptual Design Review (CDR) last year. Taking into account such concept, comprehensive transient-state thermal analyses are here explained into detail. They have been implemented in order to obtain important inputs for the mechanical design of the spectrograph.

Finally, the entire instrument includes also a visible spectrograph which can provide important feedback and complement the science developed by means of the NIR Spectrograph. However, the visible spectrograph is not being studied in the present work.

*Contact S. Becerril (santiago@iaa.es) for further information.

## 2. THERMAL REQUIREMENTS AND BOUNDARY CONDITIONS

In order to ensure that the instrument reaches 1m/s accuracy in the radial velocity measurements, the preliminary requirements used for the present study limit the thermal stability to ±0.01K (within year periods) over a working temperature of 243K. Mention has to be done of the fact that the accuracy over the absolute temperature reached on the instrument is not critical at all. The figure above mentioned is the upper limit: any part of the instrument must be cooled down under 243K, but once the thermal steady-state is reached the temperature of all the systems within the spectrograph may only range within ±0.01K, within year-scale periods. On the other hand, the detector has a specific thermal enclosure which must ensure the appropriate temperature (80K) for its operation.

The most convenient location for the CARMENES-NIR spectrograph in the 3.5m Calar Alto Telescope facilities is the Coudé Room, which is not exposed to outdoor temperature variations. The environment conditions of such location are the following:

- The air temperature in the Coudé Room is around 12ºC.
- Temperature measurements: Mid-term (3-4 months) precise measurements of the temperature in the Coudé Room have been collected. The temperatures were collected by mechanical thermometers installed at the secondary Coudé Room from mid March 2009 to the first of July 2009. Maxima and minima collected over periods of 8-9 days are plotted in Figure 1. Results are listed below:
    o Intra-day variations are within 2ºC.
    o Differences within a week can be as large as 3.5ºC.
    o Seasonal differences can be larger than 6ºC.

Although the thermal conditioning of the Coudé Room is not within the fine range, it provides very good temperature conditions as compared as other locations as the Telescope Hall. In other words, the system providing the thermal-controlled environment for the NIR spectrograph will be by far less exposed to temperature variations than in the case of other locations. This is a crucial point regarding the thermal/mechanical stability of the instrument.

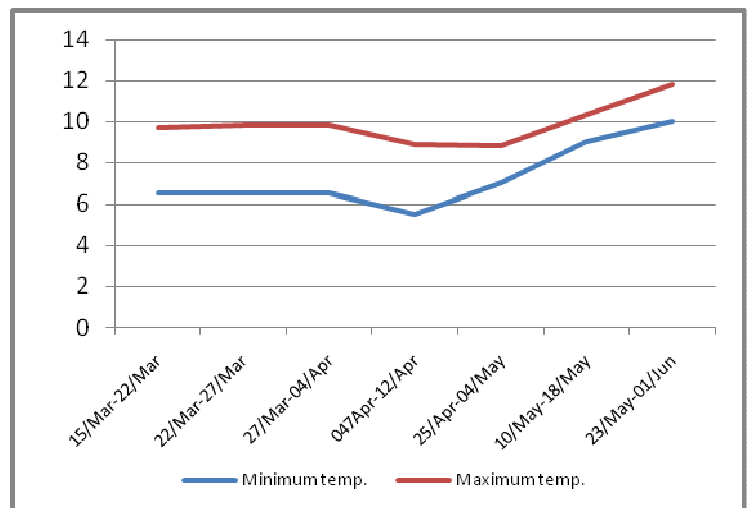

Figure 1. Maximal and minimal temperatures during the period from 15-Mar to 01-Jun 2009 of the small secondary Coude room.

## 3. NIR SPECTROGRAPH SYSTEM CONFIGURATION

Figure 2 shows the optical layout of the NIR spectrograph presented at CDR. All the systems there, except for the detector, are housed inside a common Vacuum Tank. Concerning the stability requirements, the need of a vacuum environment has been found important to avoid heat convective flows inside the spectrograph which may compromise the thermal stability. Figure 3 gives a more accurate sense about the dimensions of the spectrograph.

On the other hand, the instrument environment should be kept stable as long as possible. It applies not only to operation time but also during dead times since such stability must be kept in the very long run (years scale). Only in case of maintenance, this stability should be perturbed. This philosophy is also translated to the mechanics related to the Optical Bench and the optomechanics. That is the reason why no mechanisms are foreseen to be inside the Vacuum Vessel. In addition, the vacuum pumps must be switched off during observation.

Since the working temperature (243K) is not far below the ambient temperature, the approach chosen to properly cool the instrument is based on thermal conditioning of the environment. In addition, this solution has been successfully implemented in the HARPS[1] instrument. Nevertheless, the CARMENES case does not match the same working temperature as the HARPS instrument since the latter works at 290K, which implies an important difference in terms of heat flow. This solution leads to a lower degree of complexity in terms of optomechanical design as compared to cryogenic instruments.

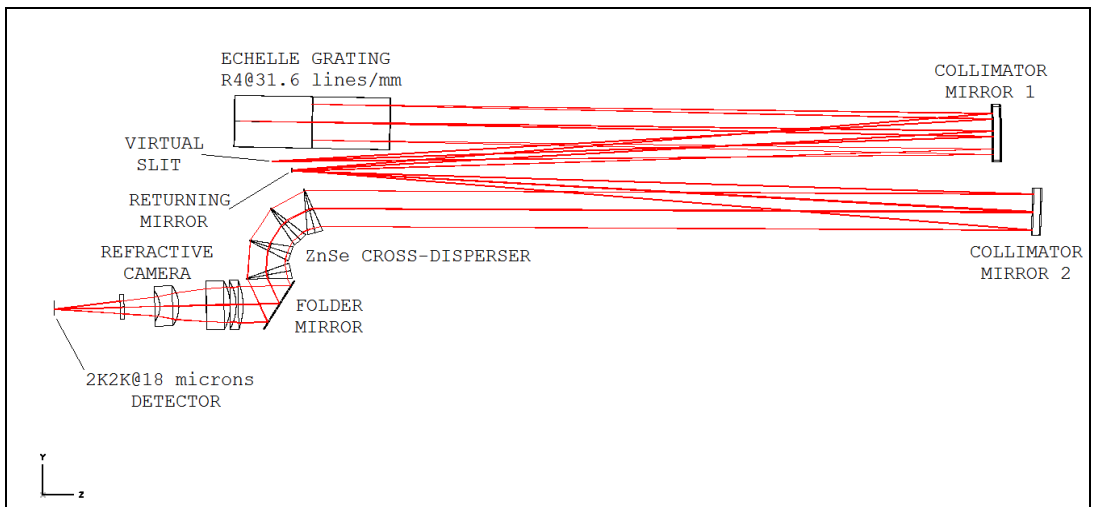

Figure 2: Basic optical layout of CARMENES-NIR.

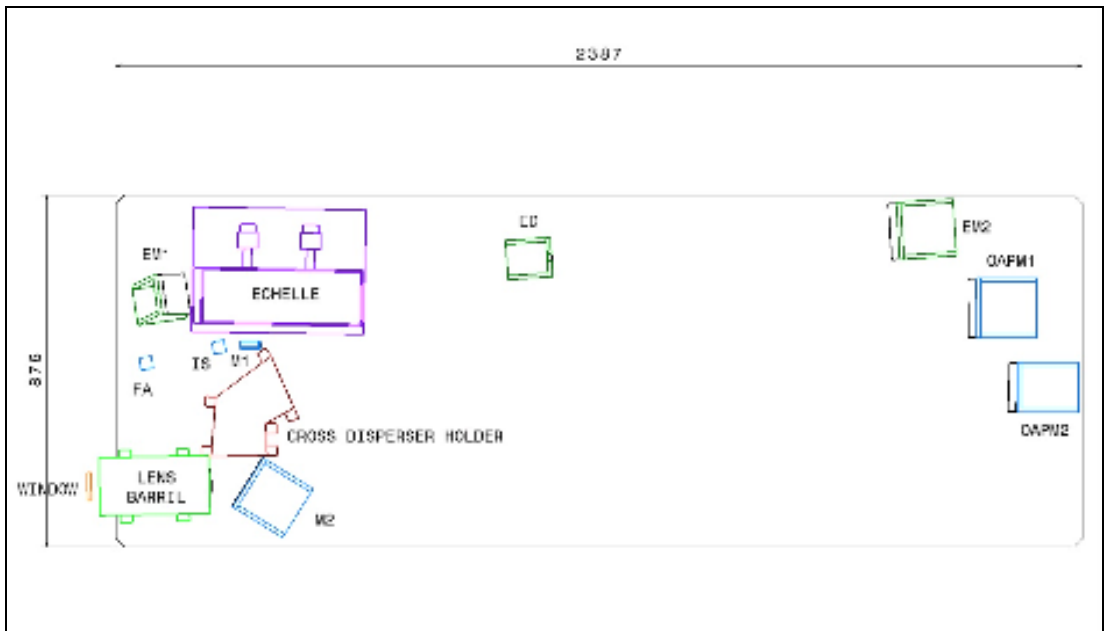

Figure 3. General layout showing the Optical Bench and the location of the optomechanical subsystems.

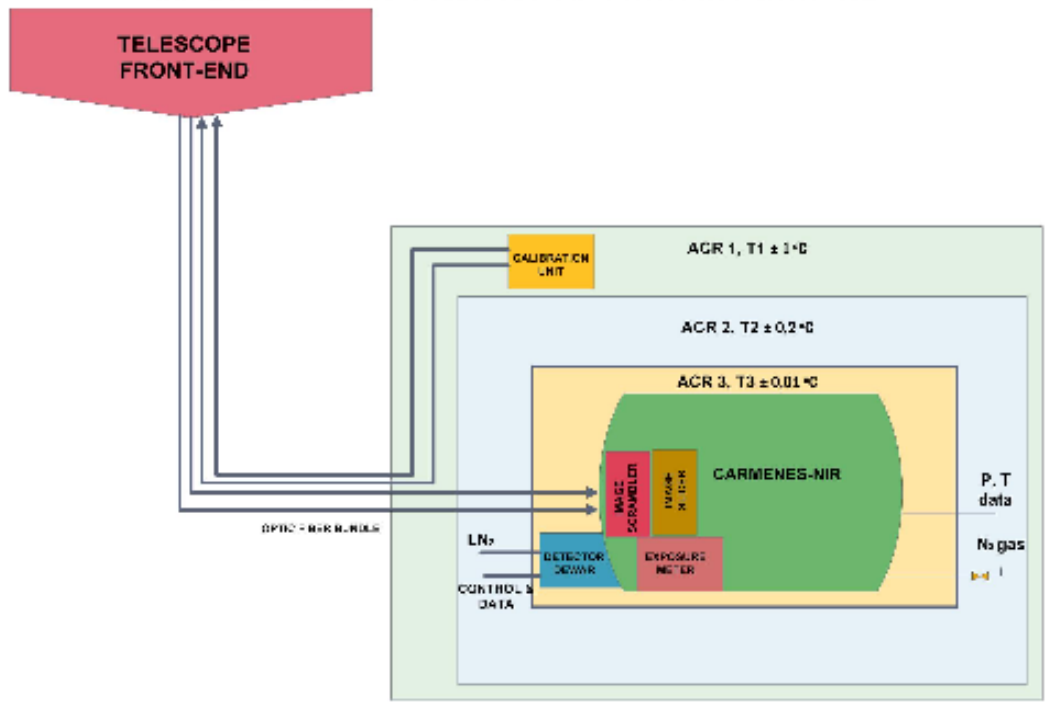

Figure 4. CARMENES NIR Block Diagram

The present concept is based (see Figure 4) on three temperature-controlled rooms (TCR) in cascade, each of them housing the immediately smaller one. The TCR1 room is the largest one and contains the intermediate room (TCR2), which, in turn, contains the TCR3 room. Finally, the latter houses the CARMENES-NIR spectrograph and provides its working temperature. Each TCR provides a thermal stage with a certain degree of stability. Thus, it is inside of the TCR3 where the working temperature is reached.

In order to better preserve the thermal stability on the IR Spectrograph, thermal losses should be minimised. That's why conduction coming from the floor must be avoided. In practice, it means that a false floor thermally isolated from the solid floor should be implemented. Similarly, since the Vacuum Vessel will be mechanically attached to the solid floor, some isolating pads should be implemented in order to minimize thermal conduction losses.

## 4. THERMAL ANALYSES

The thermal analyses here performed have been focused on two main aspects:

- Thermal stability: This critical requirement has to be fulfilled within ±0.01K.
- Thermal transient stage: This issue does not apply during the normal working of the CARMENES NIR spectrograph but it does during Assembly/Integration/Verification (AIV) phase, since a certain number of vacuum-and-cooldown cycles shall be necessary to provide the proper alignment and setup to the optomechanics. Therefore, even if the AIV phase is going to take place only once in the instrument lifetime, extremely long transient stages are not acceptable. Typically, these stages should not be longer than 3 – 4 days.

## 4.1. Thermal stability analyses

In section 3, the concept of the thermal system for the instrument to get the working temperature has been described. Nevertheless, how the thermal stability is achieved is not yet solved. Next, two different alternatives to ensure the stability requirement have been taken into account:

- Alternative 1: Thermal "inertia" of the system
  The environment temperature is not controlled to achieve the ±0.01K requirement, but to reach ±0.1K. A coarser degree of thermal control is then enough for TCR3. Thermal stability (±0.01K) in the spectrograph is wanted to be achieved by large "inertia" of heat transfer procedures. Concerning radiation heat transfer issues, such guideline would be achieved by inserting many radiation shields between the Vacuum Vessel and the Optical Bench. Typically, that should be implemented through a high degree of radiation shielding (Multilayer insulation (MLI), very low emissivity values).
  The benefit of the present alternative would lie on a lower degree of complexity for the temperature control of the TCRs. Eventually, the enclosure providing the last stage of thermal conditioning (TCR3) could be saved, thus also providing noticeable decrease of the cost. On the other hand, transient times in the AIV may exceed the reasonable limit of 3-4 days for each cooldown/vacuum cycle.

- Alternative 2: Environment thermal control
  The environment temperature is controlled to achieve both the working temperature and the thermal stability requirement. It requires a more complex temperature control for TCR3 than in the case of the Alternative 1. If the environment fulfils the thermal requirements and the conduction losses are minimised, all the hardware inside TCR3 also will match ±0.01K stability.

  Such approach leads to reverse strategy in terms of thermal "inertiae" of the subsystems involved, which would produce some benefits in terms of transient stage time. According to specialized feedback on that domain, the stability requirement of 0.01K is feasible. Furthermore, the present alternative has been already implemented in the HARPS instrument. Additionally, the fact that the location of the NIR spectrograph is not subjected to outdoor temperature variations is a crucial point on this issue. Indeed, the variation of the heat flow to TCR1 is this way noticeably limited.

  This alternative would lead to designs focused, from the thermal point of view, on producing short cooldown transient times. Unlike the Alternative 1, in principle, there is no point here to provide high inertia of the thermal transfer procedures. Therefore, no radiation shielding is foreseen here. Furthermore, mass should be intended to be minimized in order to get shorter cooldown transient times. This way, radiation heat transfer from Optical Bench to Vacuum Vessel would be promoted.

The present analysis has been carried out to estimate the time required for the Optical Bench to change its temperature by 0.01K provided that the TCR3 is exposed to a temperature step variation of 0.1K. In particular, the present analysis starts when the temperature inside TCR3 is 243.1K and gives the time elapsed for the Optical Bench to reach 243.01K. This applies only to Alternative 1, which, in turn, has been studied according to different cases (depending on the degree of radiation shielding between the Vacuum Vessel and the Optical Bench):

- Case I: The Optical Bench "sees" directly the Vacuum Vessel. No radiation shielding is implemented.
- Case II: One Radiation Shield is considered between the Optical Bench and the Vacuum Vessel.
- Case III: In addition to the metallic Radiation Shield from Case II, 50-layers MLI insulation is considered between the Vacuum Vessel and the Radiation Shield. Consequently, radiation power transferred from the Vacuum Vessel to the Optical Bench is this way reduced by a factor 1/51 as regard to the Case I. Obviously, this configuration is the one with highest "inertia" in terms of heat radiation transfer.

### 4.1.1. Assumptions

The assumptions which have been considered are the following:

- The key components involved in the present analysis are the Vacuum Vessel (VV), the Radiation Shield (RSh) and the Optical Bench (OB)
- In terms of geometry, some simplifications have been also implemented for this first-order analysis. The Vacuum Vessel, the Radiation Shield and the Optical Bench have been considered as cylinder-shaped envelopes. Obviously, it is mainly the Optical Bench which is far away from such assumption. The error so produced concerns the view factor of the Vacuum Vessel (or Radiation Shielding) as regard to the Optical Bench. Nevertheless, the analysis results are not especially sensible to this parameter.
- The dimensions taken into account are the following:

$$D_{VV} = 1.2m;\ L_{VV} = 2.6m;\ D_{RSh} = 1.1m;\ L_{RSh} = 2.5m;\ D_{OB} = 0.9m;\ L_{OB} = 2.4m$$

Where $D_{VV}$, $D_{RSh}$ and $D_{OB}$ are the respective diameters of the VV, the RSh and the OB and $L_{VV}$, $L_{RSh}$ and $L_{OB}$ are the respective lengths pf the VV, the RSh and the OB.
- The Vacuum Vessel is considered to be made of stainless steel. Both the Radiation Shield and the optical Bench are considered to be made of aluminium. Masses considered for the present study are 1000kg, 85kg and 400kg for the VV, the RSh and the OB, respectively.
- Isothermal surfaces are considered. This condition is closer to be fulfilled in case of high conductivity materials and minimization of conduction losses.
- Surfaces are considered diffuse. For electropolished metal surfaces, emissivity can be considered around 0.03-0.04.
- Conductions losses are neglected.
- The main items involved to this analysis are the Vacuum Tank, the Radiation Shielding and the surrounding environment.
- Thermal gradients along the wall thickness of the Vacuum Vessel and the Radiation Shielding are negligible, due to the high conductivity of the materials involved.
- The natural convection coefficient ($h_{nat}$) between the air inside the TCR3 and the VV has been estimated 5W/m²K.
- Times elapsed to change the temperature of the air inside TCR3 are considered negligible. In fact, convection heat transfer power is so high as compared to radiation process that the times elapsed to change the temperature of the VV might be also neglected.

**4.1.2. Analysis methodology**

The analysis methodology here explained is based on the Case I. The same methodology has been used for the rest of cases. So, the equations applicable are the same, the main difference lying on the fact that the Floating Shields (or MLI) are the components involved in the radiative heat transfer procedures with both the Vacuum Vessel and the Optical Bench. The present methodology has been based on a discrete series of intermediate states, a regular temperature step *ΔT* being applied from one state to the next one:

1. The routine starts from the moment in which the temperature inside the TCR3 Vacuum Vessel has changed to 243.1K, the rest of key temperatures still being 243K.

2. A step temperature variation (*ΔT*) is set to the Vacuum Vessel. The smaller the temperature step is, the more accurate the results are. Typically, since the stability requirement is 0.01K, the step variation has to be set as an order of magnitude lower (0.001K). For better accuracy, an additional order of magnitude is decreased, the step variation being 0.0001K.

Then, the heat transfer power is found from the TCR3 ambient (convection) to the Vacuum Vessel and from the latter (radiation) to the Optical Bench through the following equations:

$$(W_{TCR3 \to VV})_0 = h_{nat} \cdot A_{VV} \cdot ((T_{TCR3})_0 - (T_{VV})_0) \qquad (1)$$

where $A_{VV}$ is the area of the Vacuum Vessel involved in convection heat transfer, $h_{nat}$ is the natural convection coefficient and subindex $0$ means the step number 0 of the present routine. $(T_{TCR3})_0$ and $(T_{VV})_0$ are the temperatures of the inside of the TCR3 and the Vacuum Vessel at t=0, respectively.

$$(W_{VV \to OB})_0 = \frac{\sigma \cdot \left((T_{VV}^4)_0 - (T_{OB}^4)_0\right)}{\frac{1-\varepsilon_{VV}}{\varepsilon_{VV} \cdot A_{VV}} + \frac{1}{F_{VV \to OB} \cdot A_{VV}} + \frac{1-\varepsilon_{OB}}{\varepsilon_{OB} \cdot A_{OB}}} \quad (2)$$

where $\varepsilon_{VV}$ and $\varepsilon_{OB}$ are the emissivity coefficients of the surfaces of both the VV and the OB. $F_{VV \to OB}$ is the view factor of the OB from the VV.

3. The routine finds $t_1$ as the time elapsed for the temperature of the Vacuum Vessel to change by 0.0001K according the net heat power incoming $(W_{VV})_{net}$ and the energy required to change by 0.0001K the temperature of the Vacuum Vessel:

$$((W_{VV})_{net})_0 = (W_{ACR3 \to VV})_0 + (W_{VV \to OB})_0 \quad (3)$$

$$(Q_{VV})_1 = m_{VV} \cdot (C_p)_{VV} \cdot ((T_{VV})_1 - (T_{VV})_0) \quad (4)$$

where $m_{VV}$ is the mass of the VV, $(C_p)_{VV}$ is the heat capacity of the VV and the sub-index $1$ means the step number 1.

$$t_1 = \frac{(Q_{VV})_1}{((W_{VV})_{net})_0} \quad (5)$$

4. The temperature of the Optical Bench after $t_1$ seconds is found as follows:

$$(T_{OB})_1 = \frac{(W_{VV \to OB})_1 \cdot t_1}{(C_p)_{OB} \cdot m_{OB}} + (T_{OB})_0 \quad (6)$$

5. Now, a new temperature step of 0.0001K is applied to the Vacuum Vessel. This procedure is iterated (through temperature steps of 0.0001K) until the temperature of the Optical Bench is 243.01K.
After $i$ steps, the temperature of the Vacuum Vessel $(T_{VV})_i$ is equal to:

$$(T_{VV})_i = 243 + i \cdot \Delta T \quad (7)$$

and the time elapsed $(t_{TOT})_i$ up to the temperature step number $i$ on the Vacuum Vessel is:

$$(t_{TOT})_i = \sum_i t_i \quad (8)$$

where $t_i$ is the time required for the Optical Bench to change its temperature from $(T_{OB})_{i-1}$ to $(T_{OB})_i$.

The sequence of temperature variation steps (applied to a certain component) driving the routine is called "stepping sequence". In this case, the present routine arrived to some saturation because the temperature variation of the Vacuum Vessel was much faster than the one of the Optical Bench. This is so because the convective heat exchange between the TCR3 environment and the Vacuum Vessel is much more powerful than the radiative heat exchange between the latter and the Optical Bench. The term "saturation" means in this context that at some step the routine arrives to an absurd result (e.g. temperature of Vacuum Vessel higher than TCR3 temperature, reversal of the heat exchange between the VV and Optical Bench,...). When that occurs, the routine finishes the current stepping sequence and starts a new one where the temperature step is applied to another component. In Case I, a second sequence was applied to the Optical Bench, thus arriving to the target temperature (243.01K) for the latter component.

6. Once the temperature of the Optical Bench is 243.01K, $t_{TOT}$ is found.

As mentioned before, the methodology above detailed is also applicable in cases II and III. In order to simplify the analysis in Case III, the input radiation power to the Radiation Shield has been considered as 1/51 times (50-layers MLI) –as in the steady-state- the radiation power considered in Case II. The effect of the MLI is so included in the analysis. Such approximation is, in principle, not strictly correct since only the steady-state fulfils that the input radiation power to a given radiation shielding layer is the same as the output radiation power from it. Nevertheless, the analysis shows that the Radiation Shielding takes extremely long time (0.1mK/h). In practice, it means that each analysis step can be considered as a pseudo-steady state, which is valid for the present first-order analysis.

### 4.1.3. Results

Results obtained from the stability thermal analysis applied to the cases mentioned in section 4.1 are listed below:

- Case I (Figure 5): Given a 0.1K temperature step on the Vacuum Vessel, the Optical Bench temperature requires about 20.8 hours to change from 243K to 243.01K.
- Case II (Figure 6): Given a 0.1K temperature step on the Vacuum Vessel, the Optical Bench temperature requires about 57h (2.37 days) to change from 243K to 243.01K.
- Case III (Figure 7): Given a 0.1K temperature step on the Vacuum Vessel, the Optical Bench temperature requires about 1062 hours (44.2 days) to change from 243K to 243.01K.

Results so obtained for Case I and Case II are not acceptable at all according the long term requirement applicable to the thermal stability. Therefore, cases I and II are not being involved in further analyses. Case III behaves clearly better although does not fit the latter requirement either. However, it is left for further analyses since it might be compatible with eventual regular-basis calibration protocols to be applied to the instrument. Indeed, if any sort of calibration procedure was to be regularly implemented, it may turn into shorter runs applicable to the stability requirements. Currently, the CARMENES team is very involved in studies about the best solution in terms of calibration.

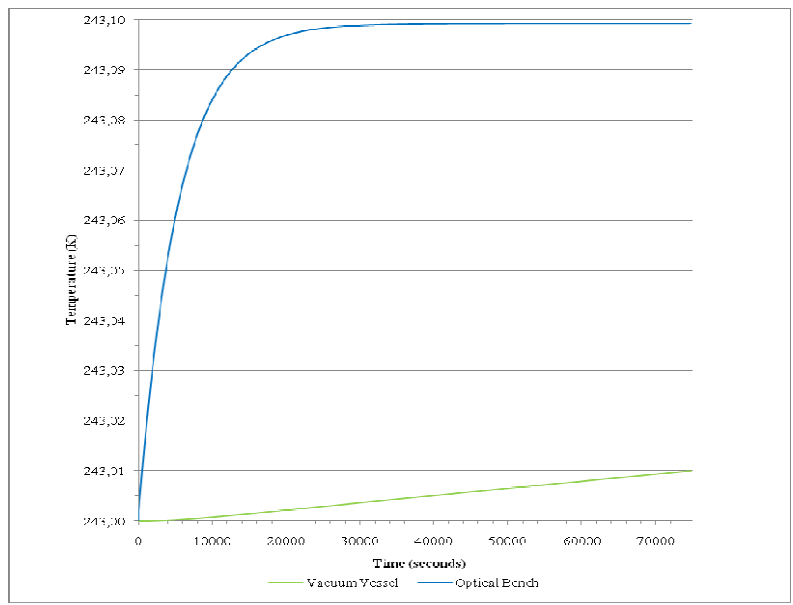

Figure 5. Case I: Thermal transient-state of the VV and the OB produced by a temperature change of 0.1K inside TCR3. The graph is limited to the time required by the OB to change the temperature by 0.01K.

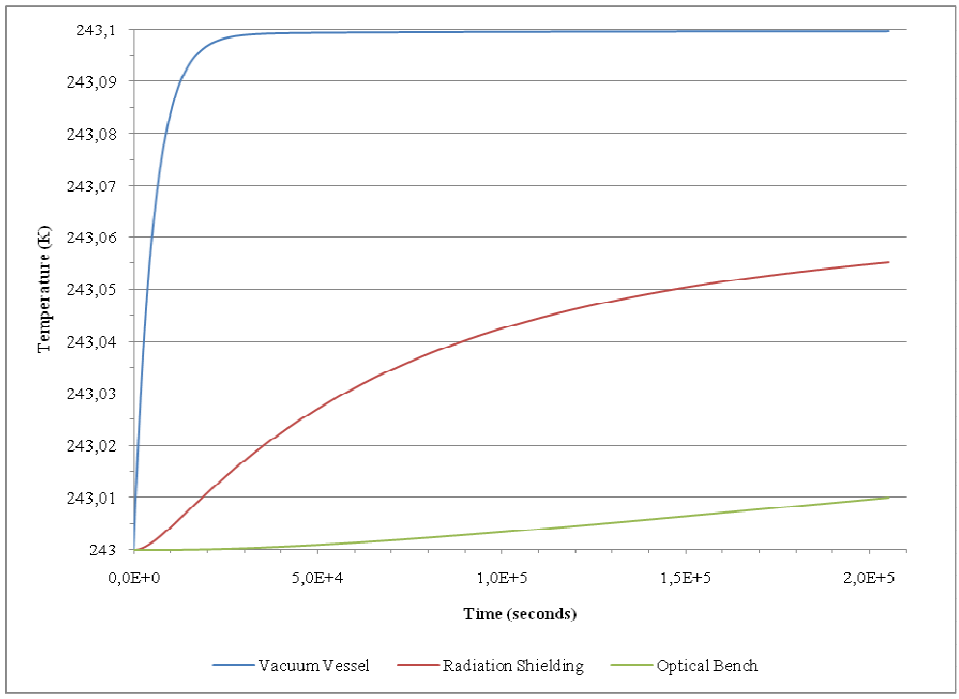

Figure 6. Case II: Thermal transient-state of the VV, the RSh and the OB produced by a temperature change of 0.1K inside TCR3. The graph is limited to the time required by the OB to change the temperature by 0.01K.

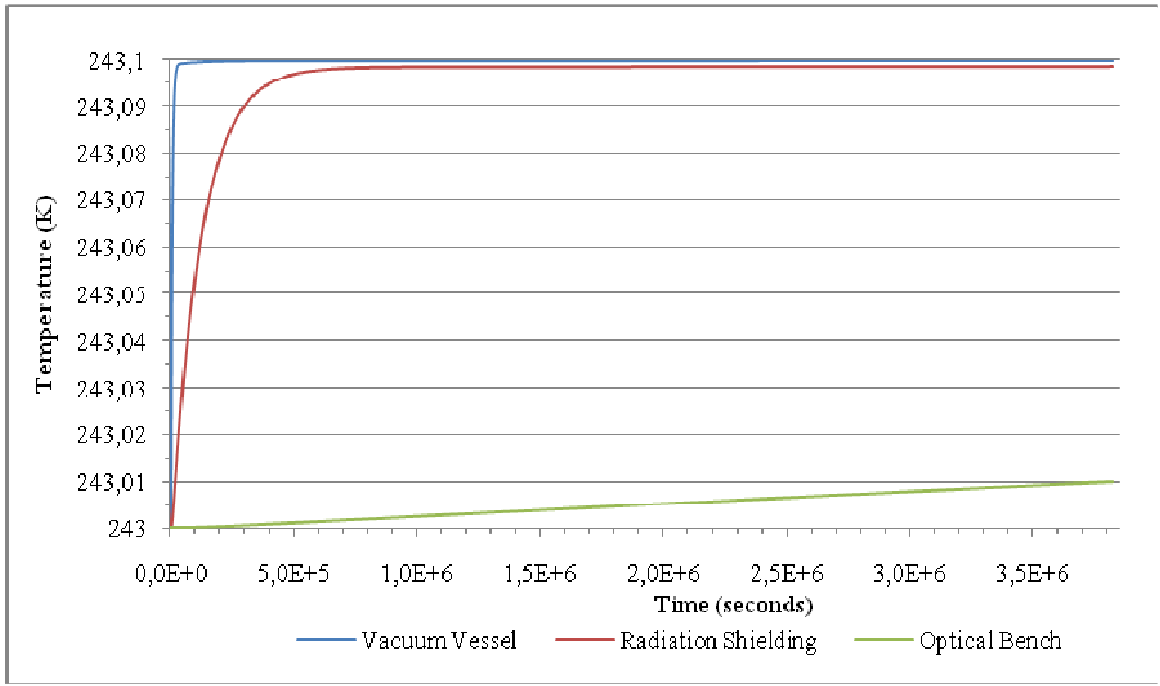

Figure 7. Case III: Thermal transient-state of the VV, the RSh and the OB produced by a temperature change of 0.1K inside TCR3. The graph is limited to the time required by the OB to change the temperature by 0.01K. The Radiation Shielding "sees" the VV and the outermost layer of the MLI.

## 4.2. Cooldown transient analysis

From results shown in section 4.1.3, only the Case III has been left for being submitted to the present analysis. Indeed, apart from the stability requirement, other important issue is the time required for the system to achieve the working temperature starting from ambient temperature. Nevertheless, it is the Case II which has been considered due to analysis simplification reasons. Therefore, one Radiation Shield has been taken into account (no MLI insulation).

The present thermal analysis is then applicable to both Alternative 1 (Case II) and Alternative 2. The latter alternative is studied according the Case I. Indeed, if the TCR3 is controlled within ±0.01K, there is no point to promote thermal "inertia" of the system by including Radiation Shielding. Therefore, only the VV and the OB are involved in Alternative 2.

The aim of this analysis consists of finding the time required for the alternatives above mentioned to reach the working temperature (within ±0.01K) from ambient temperature (293K). This objective is considered to be accomplished when all the key components (Vacuum Vessel, Radiation Shielding and Optical Bench) are at a temperature below 243K, the differences between each other being lower than 0.01K.

### 4.2.1. Assumptions

Beside the assumptions already mentioned in section 4.1.1, those which are particular for the present analysis are listed below:

- The TCR3 is supposed to be active-controlled such a way that a cooldown protocol can be implemented so that the OB cooldown time is minimized.
- According specialized feedback, one air-conditioning cooling stage can reach temperatures as low as 226K. Below this value, a second air-conditioning stage would be necessary. Thus, in the present study the TCR3 is supposed to be cooled down to 226K.
- The cooling power of the air-conditioning system can be varied such a way that the convection coefficient may be noticeably changed depending on the application. For example, in steady-state conditions, the input power to keep the thermal stability should be very limited. Alternatively, in conditions of early transient-state cooldown phases, convection heat transfer may be maximized through high input power.
- As it will be seen, the transient cooldown can be split into several phases, where the TCR3 temperature controlled may be adjusted to different temperature targets. For example, in early cooldown phases, the TCR3 temperature may be kept at 226K in order to maximize radiative heat transfer from the VV inwards.
- The time necessary to change the TCR3 air temperature has been neglected as regard the results obtained for each phase of the cooldown because, in general, convection heat transfer is much more powerful than radiative heat transfer. According the results shown in section 4.2.2, cooldown transient times are so long that, in general, the time required to cool the air inside TCR3 down does not represent a significant part of the total.

As mentioned in section 4.1, the strategy of Alternative I (case III) in terms of thermal isolation is the opposite as compared to the one applying in Alternative II. Therefore, in the first case, emissivity values around 0.03 (electropolished surfaces) are considered to maximize thermal isolation; in the second case, emissivity values around 0.1 (no polished surfaces) are considered to promote radiation heat transfer without dramatically increasing the outgassing load inside the VV. As a reminder, vacuum inside VV needn't to be in the very high range: $10^{-4}$ mbar is enough for the CARMENES NIR spectrograph.

### 4.2.2 Results

The analysis methodology is based on the same equations shown in section 4.1.2, the main differences lying on the start conditions of the routine, as well as on the components where the temperature step $\Delta T$ is applied at each phase.

#### 4.2.2.1 Alternative 1 (Case II)

Concerning this alternative, the routine implemented has shown that the cooldown process (from ambient temperature to working temperature) can be split into 5 phases (see Figure 8):

- PHASE I: Vacuum Vessel Cooldown. During this phase it is mainly the Vacuum Vessel which is being cooled down by convection. The Radiation Shielding and the Optical Bench temperatures decrease at much lower rate. The TCR3 temperature is set at 226K in order to minimize the time required for this phase. This phase is considered to be finished once the Vacuum Vessel temperature is as low as 230K, which occurs after 6100 seconds.
- PHASE II: Radiation Shielding Cooldown. The Vacuum Vessel temperature keeps decreasing up to values slightly over 226K. The Vacuum Vessel never gets 226K because of the radiation load coming from the Radiation Shielding. During this long phase, the temperature inside TCR3 is kept at 226K in order to maximize the cooldown rate of the RSh and the OB.. This phase is considered to be finished once the Radiation Shielding has attained a temperature enough below 243K such a way that, in the next phase, it converges up to the temperature of the OB. In this case this value is 234.48K. Obviously, this value has been set after some feedback obtained from the present routine. The duration of Phase II is around 1820000 seconds.
- PHASE III: Vacuum Vessel Warm-up. The Vacuum Vessel is heated up by convection ($T_{TCR3}$ = 258K) up to a temperature slightly below 243K. This phase's duration is only 1600 seconds approximately.
- PHASE IV: Radiation Shielding Warm-up. The TCR3 control is tuned to keep the VV at the same temperature as at the end of Phase III. The Radiation Shielding warms up from a temperature around 234.5K to the same temperature as the OB. In the present case, the final temperature of both the Radiation Shielding and the Optical Bench is 241.89K. Once this happens (duration of Phase IV: 171400 seconds), Phase V starts up.
- PHASE V: Vacuum Vessel tuning. The TCR3 temperature control is driven to slightly change the temperature of VV up to the same as Radiation Shielding and the Optical Bench. The duration of this phase is fully negligible as regard the total duration of the cooldown.

The duration of the cooldown in Alternative 1 (case II) is around $2 \times 10^6$ seconds (23.1 days).

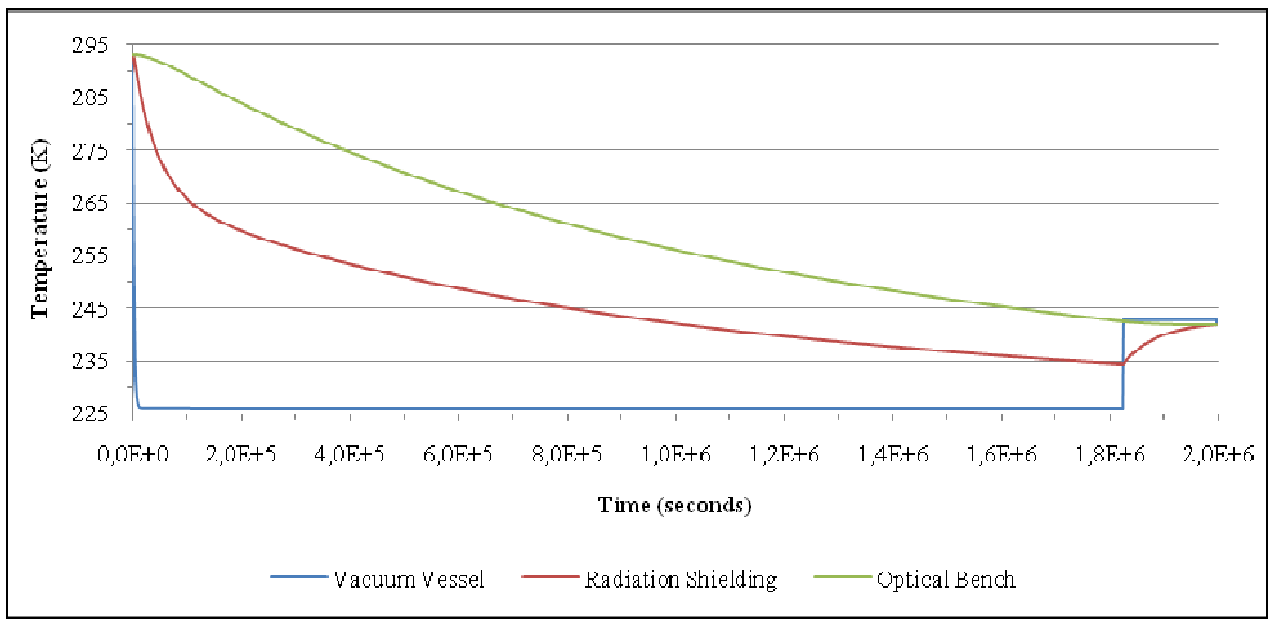

Figure 8. Alternative I – Case 2: Cooldown transient-state of the Vacuum Vessel, the Radiation Shielding and the Optical Bench from ambient temperature to working temperature. All the phases above explained can be easily found: Phase I (fast VV Cooldown), Phase II (Large period for RSh and OB Cooldowns), Phase III (fast VV Warm-up), Phase IV (RSh Warm-up converging to OB temperature), Phase V (fast tuning of VV temperature to working temperature).

### 4.2.2.2 Alternative 2

Concerning this alternative, the routine implemented has shown that the cooldown process (from ambient temperature to working temperature) can be split into 3 phases (see Figure 9):

- PHASE I: Vacuum Vessel Cooldown. During this phase it is mainly the Vacuum Vessel which is being cooled down by convection. The Optical Bench temperatures decrease at a much lower rate. The TCR3 temperature is set at 226K. The Phase I is considered to be finished once the Vacuum Vessel temperature is as low as 230K, which occurs after 2500 seconds approximately.

- PHASE II: Optical Bench Cooldown. The Vacuum Vessel temperature keeps decreasing up to values slightly over 226K. The Vacuum Vessel never gets 226K because of the radiation load coming from the Optical Bench. This phase is considered to be finished once the Optical Bench has attained a temperature slightly below the work temperature. In this case such temperature is 241.91K.

- PHASE III: Vacuum Vessel Warm-up. The Vacuum Vessel is heated up by convection ($T_{TCR3}$ = 258K) as to make $T_{VV} = T_{OB}$. The control of TCR3 temperature should be implemented on real-time in accordance with the temperature of the Optical Bench, which keeps slightly decreasing, and the inertia of both the Vacuum Vessel and TCR3 air volume. At the end, the temperature of the components is 241.89K.

The duration of the cooldown in Alternative 2 is around $2.73 \times 10^5$ seconds (3.16 days).

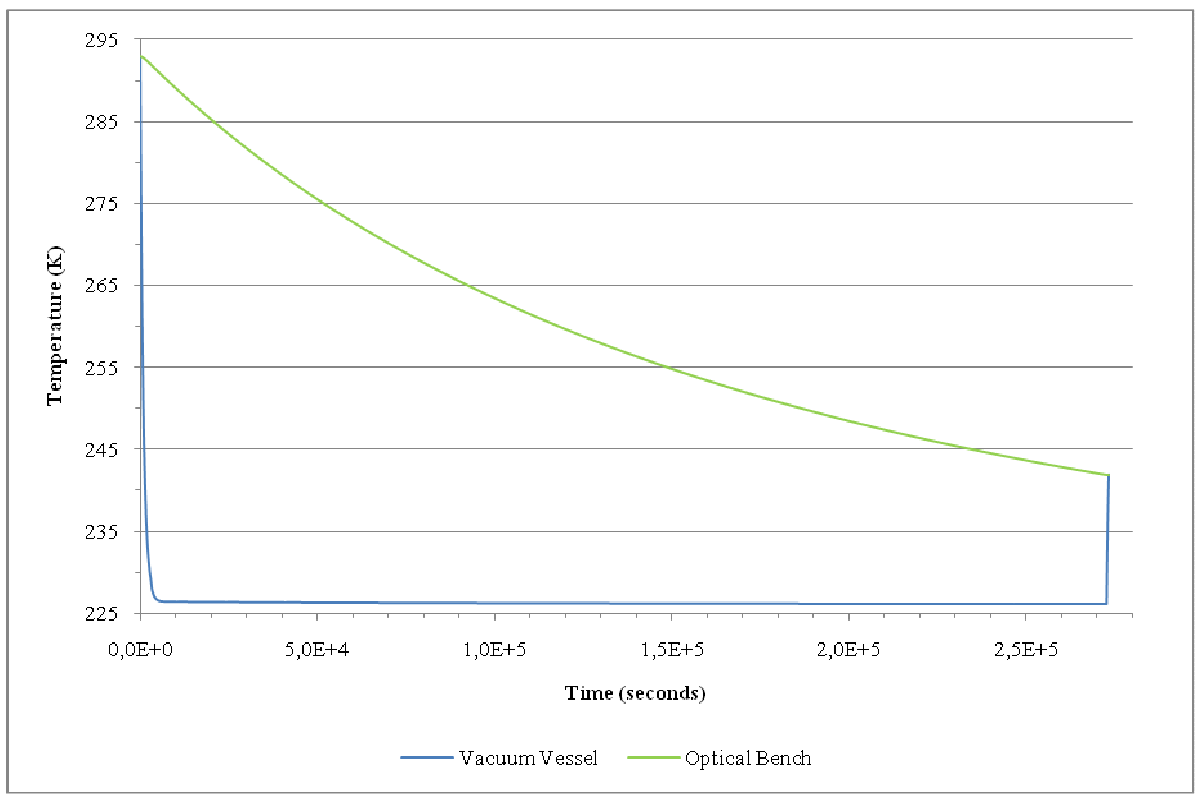

Figure 9. Case 1: Cooldown transient-state of the Vacuum Vessel and the Optical Bench from ambient temperature to working temperature. All the phases above explained can be easily found: Phase I (fast VV Cooldown), Phase II (Large period for OB Cooldown), Phase III (fast VV Warm-up to working temperature).

# 5. CONCLUSIONS AND FURTHER TASKS

Next, the conclusions extracted from the results are listed below:

- Alternative 1, even in the best case in terms of thermal stability (Case III), is not acceptable in terms of cooldown transient time. As a reminder, it is the Case II which has been submitted to the cooldown transient analysis, being the result about 23 days. Therefore, the time for the Case III must be much longer.

- At present, Alternative 2 is the approach selected for providing the thermal stability to the NIR Spectrograph, which is ensured by the proper temperature control system of TCR3, while the cooldown time from ambient temperature to working temperature is kept within reasonable limits (around 3 days long). However, this alternative does not promote thermal "inertia" (poor reflectivity factors, no Radiation Shielding), which, on the other hand, is good for stability. That is why, according the present results, a new alternative will be further analyzed:

    o This alternative would present a metal Radiation Shielding and 50-layers MLI, the latter being placed between the VV and the RSh. The OB only would "see" the Radiation Shielding, which would include a coil for cooldown. The temperature-controlled coolant circulating through the coil should provide reasonably short cooldowns while keeping the OB temperature stable within ±0.01K.

- On the other hand, comprehensive efforts are being done within the CARMENES team in order to choose the best science calibration system and set an appropriate calibration protocol. From these tasks, the period upon which the thermal stability requirement applies might be changed. This could imply new analysis on the basis of this new scenario.

- A preliminary design phase should be started up to set the hardware necessary to ensure the stability requirement. In order to implement this phase properly, a comprehensive temperature measurement campaign must be implemented in the 3.5@Calar-Alto Coudé Room. A basic parameter to be known is the response time necessary to compensate the heat flow variations through the TCRs. This is very depending on how steep are the temperature variations in time.